# Combined Emerging Capabilities for Near-Earth Objects (NEOs)

S.N. Milam (NASA/GSFC), H.B. Hammel (AURA), J. Bauer (UMD), M. Brozovic (JPL), T. Grav (PSI), B.J. Holler (STScI), C. Lisse (JHU/APL), A. Mainzer (JPL), V. Reddy (Arizona), M. E. Schwamb (Gemini Obs.), T. Spahr (NEO Sciences, LLC), C.A. Thomas (NAU), D. Woods (MIT, Lincoln Labs)

**Goal:** Assess the joint capabilities of emerging telescopes for near-Earth objects (NEOs) survey and characterization, and what they will add to the current capabilities or replace. NASA telescopes in prime mission, in development, or under study, and requested for this assessment, include:
- The Transiting Exoplanet Survey Satellite (TESS)
- The James Webb Space Telescope (JWST)
- The Wide Field Infrared Survey Telescope (WFIRST)
- The Near-Earth Object Camera (NEOCam)

Also requested for this assessment is the Large Synoptic Survey Telescope (LSST), an 8.4-meter ground-based telescope in development by the National Science Foundation and Department of Energy (DOE), with the capability to discover and catalogue NEOs.

## 1. Introduction

Near-Earth objects (NEOs) are asteroids and comets that have their perihelion distance less than 1.3 au. To date, we know about ~20,000 Near-Earth Asteroids (NEAs) and just over 100 Near-Earth Comets (NEC). Potentially Hazardous Asteroids (PHAs) are NEAs that are larger than ~140 m in diameter and that have an Earth Minimum Orbit Intersection Distance (MOID) smaller than 7.5 million km. We currently know about ~2000 PHAs. The most critical questions about PHAs are: *When is the next impact likely to occur?* and *How bad will it be?* To answer the former, we must discover PHAs in large numbers and measure their orbits with sufficient precision to understand their impact probabilities on timescales of order a century or two. The answer to the latter question depends largely on kinetic energy, which scales as $\frac{1}{2} \times$ mass $\times$ velocity$^2$, or as ~diameter$^3 \times$ density $\times$ velocity$^2$. Most survey telescopes focus on discovery of PHAs in an effort to address the first question, with little effort to provide information about the objects' impact energy beyond their absolute visible magnitude. Characterization efforts are restricted to a much smaller number of objects. For example, ~10% of known NEOs have been spectrally characterized; (Binzel et al. 2019) and ~10% of these NEOs are also PHAs

## 2. Current NEO Capabilities

**Survey:** Current dedicated NEO surveys are primarily conducted by ground-based telescopes operating at visible wavelengths with 1-2 m class apertures, with the exception of the WISE/NEOWISE mission, which operates at mid-infrared (IR) wavelengths from low-Earth orbit. Survey completeness exceeds 90% for NEAs larger than 1 km in effective spherical diameter and is estimated at ~30% for NEAs larger than 140 m (Stokes et al. 2017). However, completeness is of order 1% or less for objects of the scale of the Tunguska and Chelyabinsk impactors (~20-50 m). Table 1 summarizes the currently operating surveys.

Another facility expected to become operational within the next year include the Space Surveillance Telescope (SST; Monet et al. 2013). Technical specifications are restricted due to its primary funding source (DARPA), but its aperture is 3.5 m with a 6 sq deg FOV. The SST is currently being transported to Australia for recommissioning. Its primary mission objective is not NEO-related, requiring fast ($\leq 2$ sec) exposures, and negotiations for access to the NEO data with the sponsoring agency (U.S. Airforce) are underway.

**Table 1: Currently operating NEO surveys feature modest-sized apertures**

| Survey | Location | Start Year | Aperture | Wavelength | Approx. Limiting Mag |
|---|---|---|---|---|---|
| **Catalina Sky Survey (CSS)** | Arizona | 2005 | 1.5 m | Visible | 21.8 (vis) |
| **Catalina Sky Survey (CSS)** | Arizona | 1998 | 0.68 m | Visible | 19.5 (vis) |
| **Pan-STARRS 1 & 2 (PS1 and PS2)** | Hawai'i | 2010, 2017 | 1.8 m | Visible | 22.0 (vis) |
| **ATLAS** | Hawai'i | 2017 | 2x0.5 m | Visible | 19.5 (vis) |
| **Zwicky Transient Facility (ZTF)** | California | 2017 | 1.2 m | Visible | 20.5 (vis) |
| **WISE /NEOWISE** | Low-Earth orbit | 2010 | 0.4 m | Mid-IR | ~22-23 (~vis) |

**Characterization:** Impact energy scales as diameter$^3$ x density x velocity$^2$. While velocity comes from knowledge of orbit, impact energy depends strongly on diameter and less so on density. Thus, knowledge of diameter is critically important to determining impact energy; a factor of 2 diameter uncertainty translates to nearly an order of magnitude uncertainty in impact energy. However, efforts to understand anything beyond an object's orbit and absolute visible magnitude, H, are restricted to a much smaller sample: of the ~20,000 NEOs known today, only a few thousand have been the subject of characterization efforts. Other parameters such as rotational period, spin pole orientation, bulk density, surface properties, shape, and potential satellites will also influence mitigation strategies should an object with a high impact probability be discovered. Different observational methods yield differing parameters, summarized in Table 2.

Effective spherical diameters are most directly determined using radar or IR radiometry and can be measured to within ~10-20% with good quality data (Wright et al. 2018, Mainzer et al. 2011, Nolan et al. 2013, Tedesco et al. 2002); stellar occultations also obtain diameters, but these observations tend to be limited for NEOs (Herald et al. 2018). Direct measurements of density are known for a handful of NEOs. These are derived from in situ spacecraft visits (e.g. Itokawa; Yoshikawa et al. 2007), from radar and optical observations of the Yarkovsky effect candidates (Chesley et al., 2003; Chesley et al., 2014), and from radar observations of binary NEAs (e.g.

Margot et al., 2002). To date, radar observed 57 out of 64 known binaries in the NEA population, and at least 50% of the observed objects have data suitable for density estimates. Similar to albedo and colors, density is broadly correlated with taxonomic class.

## 3. Summary of Emerging Asset Capabilities:
## 3.1. NEO Searches

While the current surveys cannot reach 90% completeness for potentially hazardous asteroids (PHAs) ≥140m within the next few decades, emerging assets will provide some capability. It has been assumed that absolute visible magnitude H<22 mag is an accuracy proxy for objects that are ≥140 m; this assumes that all NEOs have albedos of 14%. However, Wright et al. (2017) showed that ~25% of NEOs have extremely low albedos, around 3%, and the remaining two-thirds are in a moderately dark population, around 17%. Using the size distribution of NEOs from Ivezic et al. (2002), Wright et al. (2017) demonstrated that a survey that aims to be 90% complete in H must reach that completeness for H<23 mag to ensure that 90% of the objects ≥140 m are detected.

Depending on its observing cadence, the predicted LSST completeness during its lifetime (2022-2032) is 80% for PHAs with H<22 mag (Chesley et al. 2017, Jones et al. 2018) and 71% for the PHAs ≥140m (Grav et al. 2016). NEOCam is a dedicated NEO survey optimized for high survey completeness for PHAs ≥140m that will also detect hundreds of thousands of smaller objects. NEOCam's expected completeness is 78% for PHAs ≥140m during its 5-year baseline mission, increasing to 88% if a 5-year extended mission is included (Mainzer et al. 2015). These numbers were reproduced in the NASA Report of the Near-Earth Object Science Definition Team named "Update to Determine the Feasibility of Enhancing the Search and Characterization of NEOs" (Stokes et al. 2017). This report reaffirmed that the goal of the search system should be to produce a catalog that is 90% complete for PHAs ≥140m, and found that the most cost-effective way to achieve this goal is using a half-meter IR space-based system or a combined visible and IR system located at the L1 Sun-Earth Lagrange point.

Other current or emerging assets, such as the Transiting Exoplanet Survey Satellite (TESS), the James Webb Space Telescope (JWST), and the Wide Field Infrared Survey Telescope (WFIRST), will make negligible contributions to the search effort. The JWST and WFIRST have very limited fields of view, which makes searching for PHAs nearly impossible for JWST and limited with WFIRST. While TESS has a very large field of view, its limited sensitivity and very large pixel sizes (21" per pixel) makes searching for PHAs nearly impossible.

## 3.2 NEO Characterization

The emerging assets offer some capability to characterize PHAs. TESS monitors the same 96×24 deg$^2$ field of view for 27 days, providing postage stamps of 200,000 preselected stars every 2 minutes and full frame images at 30-minute cadence. TESS will provide well-sampled rotational light curves for most of the brighter main belt asteroids and a much smaller number of PHAs, as these rarely are bright enough for TESS to detect. TESS will also provide information on thousands of stellar occultation events (Pal et al. 2018), but PHA occultations will be limited.

LSST will provide a catalog of observations of PHAs from ~0.3-1.1 μm (*ugriz*), allowing for colors to be determined for some objects. This can allow for crude determination of taxonomic classification of many PHAs using the taxonomic system of e.g. Carvano et al. (2010). However, this effort will be made significantly harder by the cadence, which results in different filters being taken days, months, or years apart, which means that effects such as rotation and phase curves can significantly affect the resulting colors and classifications. For some objects, rotation state and phase curves may be solvable.

NEOCam will provide observations in thermal IR wavelengths for most PHAs ≥140m. IR observations allow for determination of diameter (Usui et al. 2014, Mainzer et al. 2011, Tedesco et al. 2002), a key physical parameter in the impact hazard, since impact energy scales as diameter$^3$. While diameter can be determined using IR photometry alone, if visible photometry is also available, the combination of visible and IR measurements allows albedo to be determined. The combination of optical photometry from LSST and thermal IR from NEOCam provides a powerful dataset that will yield accurate diameters and albedos for the majority of PHAs ≥140m. NEOCam also includes support for a targeted mode that allows high SNR observations in two channels to be obtained, along with densely sampled lightcurves.

JWST can potentially characterize surface composition, surface alteration, and physical properties of individual PHAs using visible and near-IR (NIR; 0.7-2.5 μm) wavelengths (Thomas et al. 2016). While JWST should be able to perform sidereal tracking of a large fraction of NEOs, it is unclear how much of the observing time will be available devoted to NEOs. Additionally, observing overheads are expected to be large (~45 min) compared to the exposure time required for most NEOs (~few min). Thus, JWST will be of tremendous value for study of individual PHAs of particular interest, but will see limited usefulness in the characterization of more than a small number of objects.

## 4. Gaps in Capabilities
### 4.1. Future Surveys

The exact LSST cadence and survey region are not yet determined, but there is some chance it will extend to +10 deg ecliptic latitude (LSST Science Collaboration, 2017). LSST is in the southern hemisphere, so image quality will suffer at moderate northern ecliptic latitudes. High northern ecliptic latitudes will be entirely unobservable at faint (V>23) magnitudes. This uncovered area will reduce NEO completeness for ≥140m, and will also decrease the ability to warn of imminent impacts from small (<100m) NEOs appearing on their last approach in the northern hemisphere. No future survey currently planned will achieve high survey completeness in this smaller size scale because there are so many objects; however, enough objects will be discovered by LSST and NEOCam to enable robust statistics on the population's characteristics.

**Table 2: Characterization of NEOs is currently performed by a variety of facilities and wavelengths; however, for the vast majority of the ~19,000 NEOs known, little is known about them other than orbit and H.**

| Method & Facilities | Parameters Determined | Aperture | Approx. # of Objects | Approx. Sensitivity | Notes |
|---|---|---|---|---|---|
| **Radar**: Arecibo & Goldstone | Orbit refinement, diameter, shape, surface features, rotational state, multiplicity, radar scattering properties, masses for Yarkovsky effect candidates, constraints on optical albedos and compositions; for binaries: mass and bulk density estimates. | 305 m & 70 m | Total unique objects between Arecibo & Goldstone is ~80-110 per yr | Can get shape model for 140 m NEO at @0.06 au (Arecibo) & @0.03 au (Goldstone) | Radar detectability scales as $1/\text{distance}^4$. Full shape/rotational state modeling only possible for ~30% of detected NEOs (e.g. Nolan et al. 2013) |
| **Visible & near-IR (VNIR) colors**: Palomar, Mayall, Blanco, SOAR | Rough taxonomic class | 4-5m | ~500 | 23 V mag | Colors can provide taxonomic classification with enough wavelengths & high SNR (e.g. Carvano et al. 2010). |
| **VNIR spectroscopy**: Palomar, Mayall, Blanco, SOAR, IRTF, Gemini | Taxonomic class | 3-8m | ~2000 total (including IRTF) | 21 V mag (19.5 V for IRTF) | See e.g. Magnuson et al. (2018). |
| **Visible lightcurves**: Modestly sized telescopes | Rotational states, shapes | <1 to 2m | ~1500 | 20 V mag | Rotational state & shape from lightcurve inversion (e.g. Warner et al. 2018). |
| **Space-based thermal IR**: Spitzer & WISE/NEOWISE | Diameters, albedos, rotational states, shapes, thermal inertia | 0.4-0.85m | ~3000 | Approx. equivalent to ~22-23 V mag | Albedo requires IR + visible; thermal inertia usually requires multiple viewing geometries and/or shape information. |

**Note:** "Diameter" refers to effective spherical diameter. Both Spitzer and NEOWISE are likely to be decommissioned in the next year.

## 4.2. Future Characterization

Table 3 summarizes emerging assets that could be used for NEO characterization. Outside of the diameters provided by NEOCam and broad band colors from LSST, there is currently no guarantee of any other large-scale characterization work. In the absence of observations by an IR system, all albedos and diameters are speculative. Radar observations can obtain high-quality sizes of a few dozen objects each year, as well as put size constraints to at least several dozen more. The increasing faintness of the average NEO discovery will also result in a drop in objects characterized at optical wavelengths (e.g. colors, lightcurves, and spectra), although at least some of these objects will have future observing opportunities.

Given the limitations on rate of motion and solar elongation of both JWST and WFIRST (see Thomas et al. 2016; Holler et al. 2018), there will always be objects that cannot be observed by either telescope. Further, time is not guaranteed on either facility for this work. NEOs frequently require immediate characterization observations, and this may not be possible given scheduling and observing constraints on both solar elongation and rate of motion.

Rotation state determination requires observations across many geometries and timescales, or radar. These observations are possible for a few hundred objects per year, extending this to a larger study of hundreds or thousands of NEOs will not be possible given the available observing time. While TESS covers large areas of sky over long (~27 day) timeframes, its limiting sensitivity of V~19 means that few additional light curves beyond those currently available are likely to be obtained. Likewise, visible and near-IR (VNIR) spectroscopy for many thousands of objects will also be unlikely to occur given the competitive nature of observing proposals and the amount of time available.

NIR is essential for constraining composition, density and meteorite analogs of PHAs. While taxonomic classification is possible with broadband colors and visible spectra, it is not diagnostic of surface composition (Gaffey et al. 2002). At present, taxonomic classification using colors is limited to objects brighter than V~21 and V~19.5 for detailed surface characterization using NIR spectroscopy. NIR characterization is limited to objects whose positional uncertainties are <1 arcminute due to the limited field of view of the guide camera on the most prolific instrument used for this task (NASA IRTF). As LSST and NEOCam come online, the number of targets available for characterization each night is expected to increase exponentially, making it challenging to keep up with detailed characterization of a significant fraction of small PHAs. Additional ground- and space-based assets for characterization of NEOs would help keep characterization in pace with discovery.

A new astrophysics mission was recently selected, SPHEREx, which is 1-5 μm all-sky survey mission that will detect ~100,000 (MBA + NEO) asteroids. SPHEREx will have the sensitivity of WISE but in 100 spectral channels. NEO full spectra will be challenging if the target is moving rapidly transverse to the ecliptic, but asteroid spectra will be easily obtained.

The Goldstone and Canberra radar antennas are primarily dedicated to spacecraft tracking, and time to observe NEOs is negotiated with spaceflight projects. Currently, Goldstone devotes about 6% of its time to NEA radar observations. Demands on the 70 m antenna for spacecraft tracking have been shrinking, so more time at DSS-14 is likely to become available in the future. At Arecibo, the Environmental Protection Agency (EPA) limits radar use to ~1000 hours/year, and any increase beyond that would need to be renegotiated with the EPA.

## 5. Summary

Next-generation surveys can significantly increase the number of NEOs discovered and tracked, but the number of objects is expected to greatly outpace the ability of current facilities to determine their impact energies and physical properties. Survey facilities will provide sparse colors and lightcurves and, if thermal IR is available, diameters, albedos, and thermal inertias along with rotational states and lightcurves. For detailed taxonomic characterization via spectroscopy, future planned facilities such as JWST will offer some modest improvement in the number of objects available today. The availability of radar facilities is not expected to change dramatically in the coming decade.

**Table 3: Future facilities that could contribute to NEA characterization.**

| Asset | TESS | JWST | WFIRST | SPHEREx | NEOCam | LSST | DSN |
|---|---|---|---|---|---|---|---|
| | 4x10 cm lenses | 6.5 m mirror | 2.4 m mirror | 0.24 m mirror | 0.5 m mirror | 8.4 m mirror | Canberra radar 70 m antenna |
| | *Low-Earth Orbit* | *L2*[a] | *L2*[a] | *Low-Earth Orbit* | *L1*[a] | *Ground* | *Ground* |
| | (2018-2020)[b] | (2021-2031)[b] | (2025-2030)[b] | (2022-2025)[b] | (2024-2029)[b] | (2022-??) | (2021-??) |
| **Operational Mode** | Pre-determined survey pattern | Queue scheduled | Pre-determined survey pattern | Pre-determined survey pattern | Pre-determined survey pattern & targeted | Pre-determined survey pattern | Queue scheduled, targeted |
| **FOR**[c] | All-sky | 85-135° | 55-126° | All-sky, 75-105° | 45-120° | ~60°[d] | N/A |
| **Wavelength** | 0.6 – 1 μm | 0.6-28 μm | 0.6 -2 μm | 0.75 -5 μm | 4.6 & 8 μm | 0.3-1 μm | 4.2 cm |
| **FOV**[e] | 96°x24° | <0.1°x0.1° | 0.79°x0.43° | 11°x3.5° | 1.7°x7.1° | 3.5°x3.5° | 0.04°x0.04° |
| **Distance to 140m NEO w/ SNR=5** | Max 0.1 au[f] | Variable | Variable | | ~1.1 au at 90 deg solar elongation[g] | ~0.7-1.2 au depending on albedo[f] | <0.03 au; can detect 500 m NEA at <0.05 au |
| **Exposure Time** | 2m and 30m exposures | Variable | Variable | 30 sec exposures | 3m & 30 sec exposures | ~30 s exposures | Exposures: N/A |
| **NEA observing time** | Serendipitous observations | GO proposal opportunities | Serendipitous; GO program? | Serendipitous observations | ~22-23 hrs/day | ~9 hrs/day | ~100 hrs/year |
| **What it means for NEAs?** | Chance follow up of bright (or close) NEAs | Occasional targeted observations | Occasional targeted observations | Characterization | Systematic discovery, self-follow up | Systematic discovery, follow up | Occasional targeted observations |
| **Data products** | Astrometry | Astrometry, spectra, lightcurves, plane-of-sky images | Astrometry, spectra, lightcurves, plane-of-sky images | Astrometry, diameters, albedos, composition, lightcurves | Astrometry, diameters, albedos, thermal inertia, IR lightcurves | Astrometry, colors, sparse lightcurves | Astrometry, diameters, echo power spectra, delay-Doppler images |

NOTE: "Follow up" means that a telescope will sweep NEAs discovered by other telescopes just as a part of the regular operating routine and "self-follow up" means that a telescope will do targeted observations of the objects that it discovered. [a] Sun-Earth L1/L2 Lagrange point. [b] Primary Mission. [c] Field of Regard (Solar Avoidance). [d] Low elongation regions are accessible, with degradation in sensitivity. [e] Field of View. [f] Object at opposition. [g] Little variation based on albedo (Grav et al. 2016; Mainzer et al. 2015).

**References:**

Binzel, R.P., Lantz, C., Carry, B., Vernazza, P., Rivkin, A., Dunn, T., Bus, S., Tokunaga, A., Thomas, C., DeMeo, F., Burbine, T., Polishook, D., Burt, B., Moskovitz, N., Reddy, V., Kohout, T., Sanchez, J.A., Slivan, S., Morbidelli, A., Granvik, M., Hicks, M., Birlan, M. 2019, Compositional Distributions and Evolutionary Processes for the Near-Earth Object Population: Results from the MIT-Hawai'i Near-Earth Object Spectroscopic Survey (MITHNEOS), Icarus (In Press)

Carvano, J. M.; Hasselmann, P. H.; Lazzaro, D.; Mothé-Diniz, T. 2010, SDSS-based taxonomic classification and orbital distribution of main belt asteroids. Astronomy and Astrophysics, Volume 510, 43, 12 pp.

Chesley, Steven R.; Farnocchia, Davide; Nolan, Michael C.; Vokrouhlický, David; Chodas, Paul W., et al. 2014, Icarus, 235, 5

Chesley, Steven R.; Ostro, Steven J.; Vokrouhlický, David; Čapek, David; Giorgini, Jon D.; Nolan, Michael C.; Margot, Jean-Luc; Hine, Alice A.; Benner, Lance A. M.; Chamberlin, Alan B., 2003, Direct Detection of the Yarkovsky Effect by Radar Ranging to Asteroid 6489 Golevka, Science, 302, Issue 5651, pp. 1739

Gaffey M. J., Cloutis E. A., Kelley M. S., and Reed K. L. 2002, Mineralogy of asteroids. In Asteroids III (W. F. Bottke Jr. et al., eds.), pp. 83–204. Univ. of Arizona, Tucson.

Grav, T., Mainzer, A.K., Spahr, T. 2016, Modeling the Performance of the LSST in Surveying the Near-Earth Object Population, AJ, 151, 172

Herald, D., Dunham, D.W., Frappa, E., Hayamizu, T., Kerr, S., and Timerson, B. 2018, Asteroid Occultations V2.0. urn:nasa:pds:smallbodiesoccultations::2.0. NASA Planetary Data System.

Holler, B. J., Milam, S.N., Bauer, J.M., et al. 2018, Solar system science with the Wide-Field Infrared Survey Telescope, JATIS 4, 034003.

LSST Science Collaborations. 2017, https://github.com/LSSTScienceCollaborations/ObservingStrategy.

Magnuson, M., Moskovitz, N., Devogele, M., Gustafsson, A., Thirouin, A., Thomas, C., Skiff, B., Mommert, M., Polishook, D., Binzel, R.P., Christensen, E., DeMeo, F., Trilling, D., Burt, B. 2018, The Mission Accessible Near Earth Object Survey (MANOS): Spectrophotometric Characterization of Small NEOs. American Astronomical Society, DPS meeting #50, id.312.07

Mainzer, A., Grav, T., Masiero, J., Bauer, J., Wright, E., Cutri, R., McMillan, R., Cohen, M., Ressler, M., Eisenhardt, P. 2010, Thermal Model Calibration for Minor Planets Observed with Wide-field Infrared Survey Explorer/NEOWIES. ApJ 736, 100.

Margot, J. L., Nolan, M. C., Benner, L. A. M., Ostro, S., J., Jurgens, R. F., Giorgini, J. D., Slade, M. A., & Campbell, D. B. 2002, Science, 296, 1445

Naidu, S.P., Benner, L.A.M., Margot, J.-L., Busch, M.W., Taylor, P.A. 2016, Capabilities of Earth-based Radar Facilities for Near-Earth Asteroid Observations. AJ 152, 99.

Nolan, M.C., Magri, C., Howell, E.S., Benner, L.A.M., Giorgini, J.D., et al. 2013, Shape model and surface properties of the OSIRIS-REx target asteroid (101955) 1999 RQ36 from radar and lightcurve observations. Icarus 226, 629-640.

Pál, A., Molnár, L., Kiss, C. 2018, TESS in the Solar System. PASP 130, 4503

Ruprecht, J.D., Viggh, H.E.M., Varey, J., Cornell, M.E. 2018, SST asteroid serach performance 2014-2017. IEEE Aerospace Conference, 10.1109/AERO.2018.8396388.

Stokes, G., Barbee, B., Bottke, W., Buie, M., Chesley, S., Chodas, P., Evans, J., Gold, R., Grav, T., Harris, A., Jedicke, R., Mainzer, A., Mathias, D., Spahr, T., Yeomans, D. 2017, Report of the Near-Earth Object Science Definition Team: Update to Determine the Feasibility of Enhancing the Search and Characterization of NEOs. Report prepared at the request of NASA.

Tedesco, E., Noah, P., Noah, M., Price, S. 2002, The Supplemental IRAS Minor Planet Survey. AJ 123, 1056.

Thomas, C. A., Abell, P., Castillo-Rogez, J., Moskovitz, N., Mueller, M., Reddy, V., Rivkin, A., Ryan, E., Stansberry, J. 2016, Observing Near-Earth Objects with the James Webb Space Telescope. PASP 128, 8002.

Usui, F., Hasegawa, S., Ishiguro, M., Müller, T. G., Ootsubo, T. 2014, A comparative study of infrared asteroid surveys: IRAS, AKARI, and WISE. PASJ 66, 56.

Warner, B., Pravec, P., and Harris, A. P., 2018, Eds., Asteroid Lightcurve Data Base (LCDB) V2.0 urn:nasa:pds:ast-lightcurve-database::2.0. NASA Planetary Data System.

Wright, E., Mainzer, A., Masiero, J., Grav, T., Cutri, R., Bauer, J. 2018, Response to "An empirical examination of WISE/NEOWISE asteroid analysis and results". arXiv:1811.01454 submitted to Icarus.

Yoshikawa, M., Fujiwara, A., Kawaguchi, J., & Hayabusa Mission & Science Team, 2007, The nature of asteroid Itokawa revealed by Hayabusa, Near Earth Objects, our Celestial Neighbors: Opportunity and Risk, Proceedings if IAU Symposium 236. Edited by G.B. Valsecchi and D. Vokrouhlický, and A. Milani. Cambridge: Cambridge University Press, 2007., pp.401-416

# 7. Supplemental Material
## 7.1 Concerns on Resources

The Arecibo and Goldstone radars currently follow up and characterize ~100 NEOs/year. Each observatory spends ~400-600 hours/year observing NEOs assuming there are no equipment issues or maintenance periods. Radar provides very accurate measurements of line-of-sight distance and velocity, and radar delay-Doppler images directly display the size and the shape of an object. The NEOs observed by radar are usually at distances within 0.1 au, and they are about several hundred meters in diameter. The larger objects can be observed farther away, and the smaller objects need to be closer. The primary limitation of detectability is the distance because the signal-to-noise-ratios (SNRs) for radar drop according to ~1/distance$^4$. In addition, the pointing uncertainties need to be <30 arcsec for a target to fall within the center of radar's ~100 arcsec wide beam. Naidu et al. (2016) estimated that the Arecibo and Goldstone radars cover ~30% of the NEOs that are potentially detectable. Most of the targets that do not get selected for observations are very small objects, diameters <20 m, or they are newly discovered objects with large pointing uncertainties. Radar targets with very low SNR are scheduled only if they overlap some stronger targets. In rare instances, observations of strong radar targets do not get scheduled because the antenna is already scheduled for another project, or in Goldstone's case, radiation clearance is not allocated or the antenna cannot be staffed on short notice.

With the onset of the future survey facilities (NEOCam, LSST), the number of potential radar targets will increase dramatically. Many of these objects will be small, <100 m diameter NEOs, and they will have short radar observing windows. Quick scheduling will be crucial for these objects as well as the support of optical telescopes that can provide astrometry and reduce the pointing uncertainties in time for radar observations. We also expect to see an increase in the number of larger objects that are detectable. Scheduling of these will depend on the antenna availability and staffing. It is important to note that at least some of the objects that are not observed during their discovery apparitions will have future encounters with Earth and possibly better radar apparitions.

The Goldstone DSS-14 antenna is primarily used for spacecraft tracking, and time to observe NEOs is negotiated with spaceflight projects. Canberra's DSS-43 antenna will be in a similar situation as Goldstone. At Arecibo, there is a limit by the Environmental Protection Agency (EPA) on how much diesel fuel that powers generators can be used per year. The current limit is ~1000 hours/year, and any increase beyond that would need to be renegotiated with the EPA. The first order approach to cover more targets with radar will likely involve reducing the number of observing hours that is used on a single target and instead spreading the time over multiple asteroids.

Availability of additional ground-/space-based assets for spectral characterization of NEOs would help keep characterization in pace with discovery. The SpeX instrument on the NASA Infrared Telescope Facility (IRTF) remains an asset (wavelength coverage, spectral resolution, high altitude location) for NEO spectroscopy. The NIHTS instrument on the Discovery Channel Telescope has complementary capabilities and longitudinal coverage compared to NASA IRTF.

Developing SpeX-type instruments for larger 6-8 meter class telescopes would enable characterization of small/faint PHAs that would be discovered by LSST and NEOCam.